\documentclass[twocolumn,notitlepage,nofootinbib]{revtex4-2}
\usepackage{amsmath,amssymb,bm,graphicx,hyperref}
\allowdisplaybreaks[1]

\renewcommand\d{\partial}
\newcommand\grad{\bm{\nabla}}
\newcommand\+{\dagger}
\newcommand\<{\langle}
\renewcommand\>{\rangle}
\newcommand\up{\uparrow}
\newcommand\down{\downarrow}
\newcommand\eps{\varepsilon}

\renewcommand\k{{\bm{k}}}
\newcommand\p{{\bm{p}}}
\newcommand\q{{\bm{q}}}
\renewcommand\r{{\bm{r}}}
\renewcommand\v{{\bm{v}}}
\newcommand\ep{\eps_\p}
\newcommand\eq{\eps_\q}
\newcommand\E{\mathcal{E}}
\newcommand\G{\mathcal{G}}
\newcommand\J{\mathcal{J}}
\newcommand\K{\mathcal{K}}
\newcommand\N{\mathcal{N}}
\renewcommand\O{\mathcal{O}}
\renewcommand\P{\mathcal{P}}
\newcommand\Q{\mathcal{Q}}
\newcommand\T{\mathcal{T}}
\newcommand\Z{\mathbb{Z}}
\newcommand\coll{\mathrm{coll}}
\newcommand\AL{\mathrm{AL}}
\newcommand\MT{\mathrm{MT}}
\renewcommand\Pr{\mathrm{Pr}}
\newcommand\LHS{\mathrm{LHS}}
\newcommand\RHS{\mathrm{RHS}}

\let\Im\relax\DeclareMathOperator\Im{Im}

\begin{document}

\title{Microscopic derivation of the Boltzmann equation\\
for transport coefficients of resonating fermions at high temperature}

\author{Keisuke Fujii}
\author{Yusuke Nishida}
\affiliation{Department of Physics, Tokyo Institute of Technology,
Ookayama, Meguro, Tokyo 152-8551, Japan}

\date{March 2021}

\begin{abstract}
Motivated by the recently observed failure of the kinetic theory for the bulk viscosity, we in turn revisit the shear viscosity and the thermal conductivity of two-component fermions with a zero-range interaction both in two and three dimensions.
In particular, we show that their Kubo formula evaluated exactly in the high-temperature limit to the lowest order in fugacity is reduced to the linearized Boltzmann equation.
Previously, such a microscopic derivation of the latter was achieved only incompletely corresponding to the relaxation-time approximation.
Here, we complete it by resuming all contributions that are naively higher orders in fugacity but become comparable in the zero-frequency limit due to the pinch singularity, leading to a self-consistent equation for a vertex function identical to the linearized Boltzmann equation.
We then compute the shear viscosity and the thermal conductivity in the high-temperature limit for an arbitrary scattering length and find that the Prandtl number exhibits a nonmonotonic behavior slightly below the constant value in the relaxation-time approximation.
\end{abstract}

\maketitle
\tableofcontents

\section{Introduction}
The BCS-BEC crossover exhibited by two-component fermions with their scattering length varied has been subjected to comprehensive studies over the past two decades in ultracold-atom physics~\cite{Bloch:2008,Giorgini:2008,Zwerger:2012,Randeria:2014}.
In particular, the system at infinite scattering length is referred to as the unitary Fermi gas and has been highlighted because of its strong correlation and scale invariance (nonrelativistic conformality~\cite{Mehen:2000,Son:2006,Nishida:2007}).
These two unique aspects are both reflected not only in its universal thermodynamics but also in its transport properties.

Because the shear viscosity of a dilute gas tends to be small as the interaction is strengthened, its smallness serves as a measure of strong correlation.
The shear viscosity of the unitary Fermi gas was measured experimentally~\cite{Cao:2011a,Cao:2011b,Elliott:2014b,Joseph:2015} and found to be close to the conjectured quantum-mechanical lower bound~\cite{Kovtun:2005}.
On the other hand, the bulk viscosity of the unitary Fermi gas vanishes identically because of its conformality~\cite{Son:2007}.
Therefore, the bulk viscosity serves as a measure of conformality breaking and its vanishment for the unitary Fermi gas was confirmed experimentally~\cite{Elliott:2014a}.
Experimental measurements of other transport coefficients have also been performed, such as the thermal conductivity~\cite{Baird:2019} and the sound diffusivity~\cite{Patel:2020} of the unitary Fermi gas, as well as the shear and bulk viscosities in two dimensions~\cite{Vogt:2012}.

Theoretically, one common approach to compute the transport coefficients is the kinetic theory, which is founded on the quasiparticle approximation supposed to be applicable only in the weak-coupling or high-temperature limit~\cite{Massignan:2005,Bruun:2005,Braby:2010,Bruun:2012,Schafer:2012,Dusling:2013,Chafin:2013}.
As a matter of principle, the transport coefficients are to be microscopically computed with the Kubo formula.
In spite of its difficulty in general situations, it can systematically be evaluated with the quantum virial expansion, which adopts the fugacity as a small expansion parameter in the high-temperature limit~\cite{Enss:2011,Nishida:2019,Enss:2019,Hofmann:2020,Frank:2020}.
Therefore, it is possible to contrast the microscopic and kinetic theories in the high-temperature limit, where they were found to disagree for the bulk viscosity~\cite{Nishida:2019,Enss:2019,Hofmann:2020}.
Such a discrepancy was attributed to the fact that the Landau kinetic theory employed in Refs.~\cite{Dusling:2013,Chafin:2013} is not fully grounded even in the high-temperature limit because of the invalid quasiparticle approximation~\cite{Fujii:2020}.
On the other hand, the shear viscosity from the quantum virial expansion combined with the approximate resummation scheme based on the memory function formalism was found to agree with that from the Boltzmann equation but only in the relaxation-time approximation~\cite{Enss:2011,Nishida:2019,Hofmann:2020}.
Therefore, the complete correspondence between the microscopic and kinetic theories for the transport coefficients is yet to be established, which constitutes the main purpose of this paper.

To this end, we evaluate the Kubo formula for the shear viscosity and the thermal conductivity exactly in the high-temperature limit to the lowest order in fugacity.
After describing general formulations in Sec.~\ref{sec:microscopics}, we sum up all contributions that are naively higher orders in fugacity but become comparable in the zero-frequency limit due to the pinch singularity~\cite{Eliashberg:1962,Jeon:1995,Jeon:1996,Hidaka:2011}.
Consequently, a self-consistent equation for a vertex function is derived in Sec.~\ref{sec:kinetics}, which is shown to be identical to the linearized Boltzmann equation.
Then, the linearized Boltzmann equation is solved numerically in Sec.~\ref{sec:transport} to compute the shear viscosity and the thermal conductivity as well as the Prandtl number in the high-temperature limit for an arbitrary scattering length both in two and three dimensions.
Finally, this paper is summarized in Sec.~\ref{sec:summary} and some supplementary materials regarding the spectral representation of three-point functions and the Boltzmann equation for transport coefficients are presented in Appendixes~\ref{app:spectral} and \ref{app:boltzmann}, respectively.

In what follows, we will set $\hbar=k_B=1$ and implicit summations over repeated spin indices $\sigma=\ \up,\,\down$ are assumed throughout this paper.
The bosonic and fermionic frequencies in the Matsubara formalism are denoted by $w=2\pi n/\beta$ and $v=2\pi(n+1/2)/\beta$, respectively, for $n\in\Z$.
Also, an integration over $d$-dimensional wave vector or momentum is denoted by $\int_\p\equiv\int\!d\p/(2\pi)^d$ for the sake of brevity.

\section{Microscopics}\label{sec:microscopics}
\subsection{Hamiltonian}
We consider two-component fermions with a zero-range interaction in $d$ spatial dimensions described by
\begin{align}\label{eq:hamiltonian}
\hat{H} &= \int\!d\r\,\hat\psi_\sigma^\+(\r)
\left(-\frac\Delta{2m}-\mu\right)\hat\psi_\sigma(\r) \notag\\
&\quad + \frac{g}{2}\int\!d\r\,\hat\psi_\sigma^\+(\r)\hat\psi_{\sigma'}^\+(\r)
\hat\psi_{\sigma'}(\r)\hat\psi_\sigma(\r).
\end{align}
We work with the Matsubara formalism and the bare fermion propagator in the Fourier space is denoted by
\begin{align}
G(iv,\p) = \frac1{iv-\ep+\mu}
\end{align}
and the full fermion propagator by
\begin{align}
\G(iv,\p) = \frac1{iv-\ep+\mu-\Sigma(iv,\p)},
\end{align}
where $\ep=\p^2/(2m)$ is the energy of a single particle and $\Sigma(iv,\p)$ is the fermion self-energy.

\begin{figure}[t]
\includegraphics[width=\columnwidth]{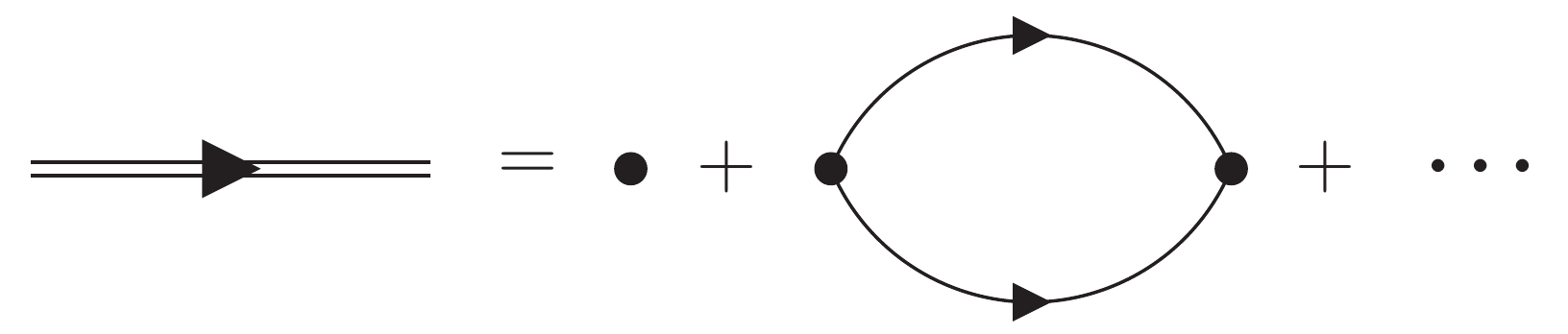}\bigskip\\
\includegraphics[width=0.7\columnwidth]{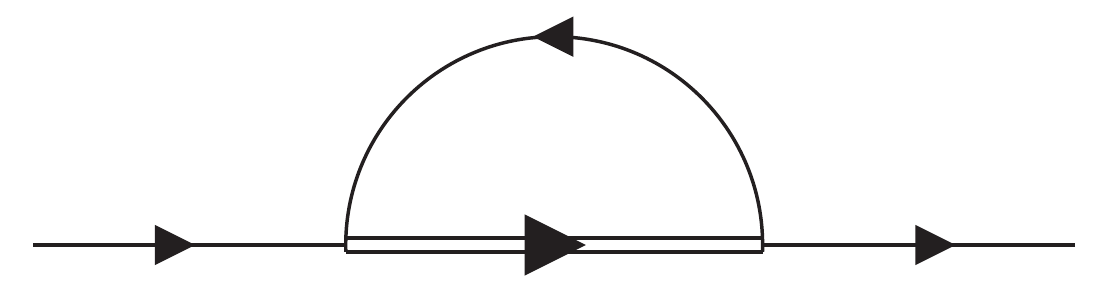}
\caption{\label{fig:self-energy}
Diagrammatic representations of the pair propagator (top) and the fermion self-energy (bottom) in Eq.~(\ref{eq:self-energy}).
The thin single and double lines represent the bare fermion and pair propagators, respectively, whereas the dot denotes a bare coupling constant.}
\end{figure}

In the high-temperature limit where the fugacity $z=e^{\beta\mu}\sim\N/T^{d/2}\ll1$ serves as a small expansion parameter~\cite{Liu:2013}, the fermion self-energy to its lowest order is evaluated as
\begin{align}\label{eq:self-energy}
\Sigma(iv,\p) = z\int_\q e^{-\beta\eq}D(iv+\eq-\mu,\p+\q) + O(z^2),
\end{align}
whose diagrammatic representation is depicted in Fig.~\ref{fig:self-energy}.
Here,
\begin{align}
D(iw,\p) = \frac{\Omega_{d-1}}{m}\frac{d-2}{a^{2-d}-[-m(iw-\ep/2+2\mu)]^{d/2-1}}
\end{align}
is the pair propagator in the vacuum, $\Omega_{d-1}\equiv(4\pi)^{d/2}/[2\Gamma(2-d/2)]=2,\,2\pi,\,4\pi$ coincides with the surface area of the unit $(d-1)$-sphere for $d=1,\,2,\,3$, and the scattering length $a$ is introduced via
\begin{align}
g = \frac{\Omega_{d-1}}{m}
\frac{d-2}{a^{2-d}-\Lambda^{d-2}/[\Gamma(d/2)\Gamma(2-d/2)]}
\end{align}
in the cutoff regularization~\cite{Fujii:2020}.

For later use, we note that the above pair propagator is simply the two-body scattering $T$-matrix and the transition rate from initial $(\p',\q')$ to final momenta $(\p,\q)$ is provided by
\begin{align}\label{eq:transition}
& W(\p,\q|\p',\q') = |D(\ep+\eq-2\mu+i0^+,\p+\q)|^2 \notag\\
& \times (2\pi)^{d+1}\delta(\ep+\eq-\eps_{\p'}-\eps_{\q'})\delta(\p+\q-\p'-\q').
\end{align}
It is also related to the imaginary part of the pair propagator via the optical theorem:
\begin{align}\label{eq:optical}
& -2\Im[D(\ep+\eq-2\mu+i0^+,\p+\q)] \notag\\
&= \int_{\p',\q'}W(\p,\q|\p',\q').
\end{align}

\subsection{Kubo formula}
According to the linear-response theory, the transport coefficients are microscopically provided by the Kubo formula~\cite{Kubo:1957a,Kubo:1957b},
\begin{align}
\eta = \lim_{\omega\to0}\frac{\Im[\chi_{\Pi_{xy}}(\omega+i0^+)]}{\omega}
\end{align}
for the shear viscosity and
\begin{align}
T\kappa = \lim_{\omega\to0}\frac{\Im[\chi_{Q_x}(\omega+i0^+)]}{\omega}
\end{align}
for the thermal conductivity~\cite{Mori:1962,Kadanoff:1963,Luttinger:1964}.
Here, $\chi_\O(\omega+i0^+)$ is a retarded correlation function at zero wave vector for an operator $\hat\O$, which is most conveniently obtained from the corresponding imaginary-time-ordered correlation function,
\begin{align}\label{eq:correlation}
\chi_\O(iw) = \frac1{L^d}\int_0^\beta\!d\tau\,e^{iw\tau}\<\T\,\hat\O(\tau)\hat\O(0)\>,
\end{align}
with an analytic continuation of $iw\to\omega+i0^+$~\cite{Altland-Simons}.

For our system described by Eq.~(\ref{eq:hamiltonian}), the off-diagonal stress tensor operator is found to be
\begin{align}
\hat\Pi_{ij} = \int\!d\r\,\frac{\d_i\hat\psi_\sigma^\+(\r)\d_j\hat\psi_\sigma(\r)
+ \d_j\hat\psi_\sigma^\+(\r)\d_i\hat\psi_\sigma(\r)}{2m} \quad (i\neq j)
\end{align}
and the heat flux operator to be
\begin{align}
& \hat{Q}_i = \int\!d\r\Biggl[-\frac{\Delta\hat\psi_\sigma^\+(\r)\d_i\hat\psi_\sigma(\r)
- \d_i\hat\psi_\sigma^\+(\r)\Delta\hat\psi_\sigma(\r)}{4im^2} \notag\\
& - \frac{\E+\P}{\N}\frac{\hat\psi_\sigma^\+(\r)\d_i\hat\psi_\sigma(\r)
- \d_i\hat\psi_\sigma^\+(\r)\hat\psi_\sigma(\r)}{2im} \notag\\
& + g\,\hat\psi_\sigma^\+(\r)\frac{\hat\psi_{\sigma'}^\+(\r)\d_i\hat\psi_{\sigma'}(\r)
- \d_i\hat\psi_{\sigma'}^\+(\r)\hat\psi_{\sigma'}(\r)}{2im}\hat\psi_\sigma(\r)\Biggr],
\end{align}
where $\N$, $\E$, and $\P$ are the number density, the energy density, and the pressure, respectively~\cite{Fujii:2018,Frank:2020}.%
\footnote{$\text{(heat flux)}=\text{(energy flux)}-(\E+\P)/\N\times\text{(number flux)}$~\cite{Mori:1962,Kadanoff:1963,Luttinger:1964}.}
Whereas the off-diagonal stress tensor operator is simply a one-body operator, the heat flux operator consists of both one-body and two-body operators.
As we will see later, contributions of the one-body operator to the transport coefficient are partly promoted from $O(z)$ to $O(z^0)$ due to the pinch singularity, but the two-body operator is supposed to provide higher-order corrections at $O(z^2)$.
Therefore, as far as the transport coefficient to the lowest order in fugacity is concerned, it is sufficient to consider only the one-body operator in the form of
\begin{align}\label{eq:operator}
\hat\O = \sum_\p\gamma_\p\,\hat\psi_{\sigma\p}^\+\hat\psi_{\sigma\p},
\end{align}
where $\hat\psi_{\sigma\p}=L^{-d/2}\int\!d\r\,e^{-i\p\cdot\r}\hat\psi_\sigma(\r)$ is a Fourier component of the field operator.
Obviously, the bare vertex function reads $\gamma_\p=p_xp_y/m$ for the shear viscosity and $\gamma_\p=[\ep-(\E+\P)/\N]\,p_x/m$ for the thermal conductivity and the corresponding transport coefficients are collectively denoted by
\begin{align}\label{eq:kubo}
\sigma_\O \equiv \lim_{\omega\to0}\frac{\Im[\chi_\O(\omega+i0^+)]}{\omega}.
\end{align}

Here, it is worthwhile to remark that $\hat\O\sim(mg)^2\int\!d\r\,\hat\psi_\sigma^\+(\r)\hat\psi_{\sigma'}^\+(\r)\hat\psi_{\sigma'}(\r)\hat\psi_\sigma(\r)$ for the bulk viscosity is essentially a two-body operator up to conserved operators~\cite{Martinez:2017,Fujii:2018,Fujii:2020}.
Therefore, our discussion below does not apply to the bulk viscosity, so that computing its leading term at $O(z^2)$ is not reduced to the kinetic theory~\cite{Nishida:2019,Enss:2019,Hofmann:2020,Fujii:2020}.

\begin{figure}[t]
\includegraphics[width=0.5\columnwidth]{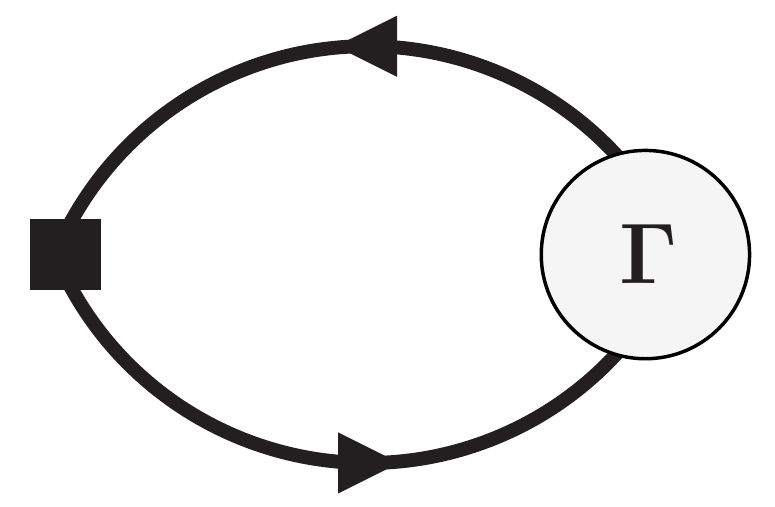}
\caption{\label{fig:correlation}
Diagrammatic representation of the imaginary-time-ordered correlation function in Eq.~(\ref{eq:time-ordered}).
The thick line represents the full fermion propagator, whereas the square and the circle denote bare and full vertex functions, respectively.}
\end{figure}

\subsection{Pinch singularity}\label{sec:pinch}
For the one-body operator in the form of Eq.~(\ref{eq:operator}), the imaginary-time-ordered correlation function in Eq.~(\ref{eq:correlation}) can formally be expressed as
\begin{align}\label{eq:time-ordered}
\chi_\O(iw) &= -\frac2\beta\sum_v\int_\p\gamma_\p\,\G(iv+iw,\p)\G(iv,\p) \notag\\
&\quad \times \Gamma(iv+iw,iv;\p),
\end{align}
whose diagrammatic representation is depicted in Fig.~\ref{fig:correlation}.
Here, the spin degeneracy accounts for the prefactor of $2$ and $\Gamma(iv+iw,iv;\p)$ is the full vertex function to be determined in Sec.~\ref{sec:vertex}.

\begin{figure}[t]
\includegraphics[width=0.8\columnwidth]{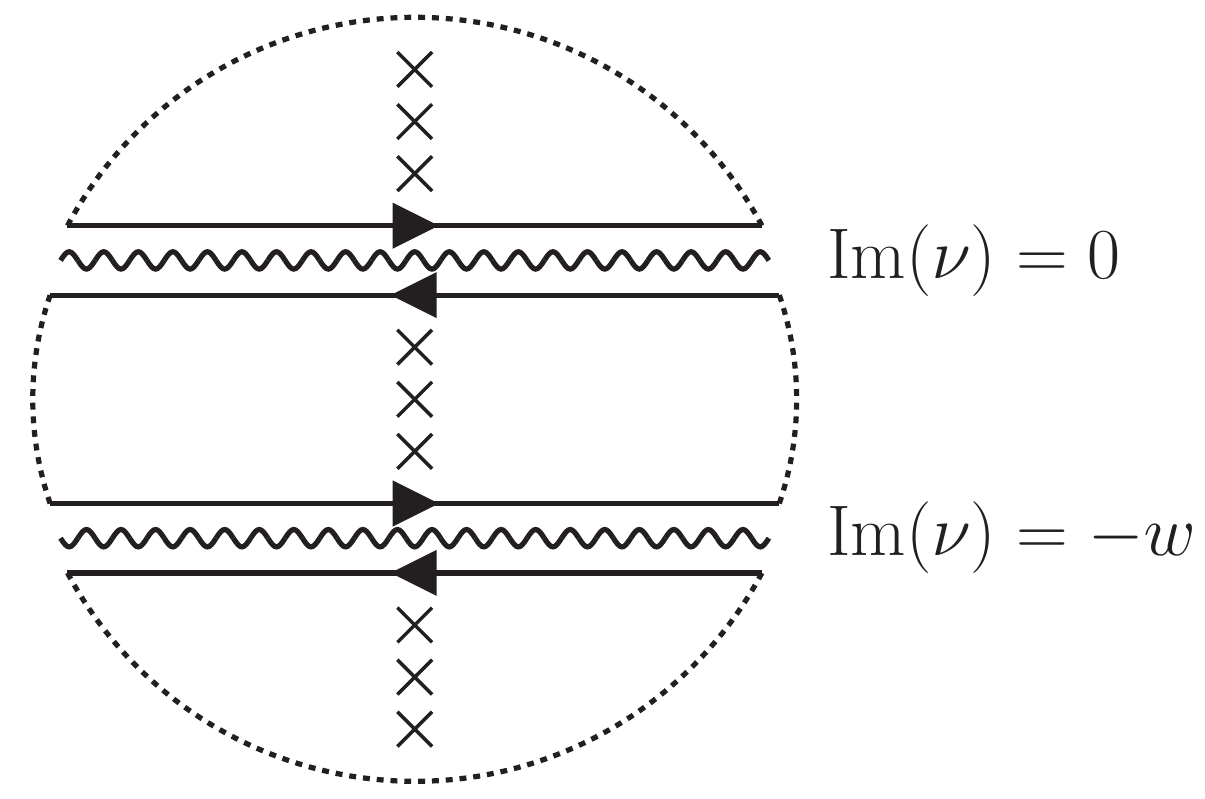}
\caption{\label{fig:contour}
Analytic structure of the integrand in Eq.~(\ref{eq:time-ordered}) with the Matsubara frequency summation replaced by the complex contour integration over $iv\to\nu$.
Besides the singularities due to the Fermi-Dirac distribution function (crosses), it may have singularities only along $\Im(\nu)=0,-w$ (wavy lines), so that the integration contour is deformed into four arrowed straight lines with vanishing contributions from infinity (dotted circle).}
\end{figure}

In order to achieve the analytic continuation to obtain the retarded correlation function, it is sufficient to know that $\G(\nu,\p)$ and $\Gamma(\nu+iw,\nu;\p)$ may have singularities only along $\Im(\nu)=0$ and $\Im(\nu)=0,-w$, respectively, in the complex plane of $\nu$~\cite{Eliashberg:1962} (see also Appendix~\ref{app:spectral}).
With the Matsubara frequency summation replaced by the complex contour integration over $iv\to\nu$, its contour is deformed into four horizontal lines as in Fig.~\ref{fig:contour}, leading to
\begin{align}
& \chi_\O(iw) = 2\int_{-\infty}^\infty\!\frac{d\nu}{2\pi i}\,f_F(\nu)\int_\p\gamma_\p \notag\\
& \times \bigl[\G(\nu+iw,\p)\G(\nu+i0^+,\p)\Gamma(\nu+iw,\nu+i0^+;\p) \notag\\
&\quad - \G(\nu+iw,\p)\G(\nu-i0^+,\p)\Gamma(\nu+iw,\nu-i0^+;\p) \notag\\
&\quad + \G(\nu+i0^+,\p)\G(\nu-iw,\p)\Gamma(\nu+i0^+,\nu-iw;\p) \notag\\
&\quad - \G(\nu-i0^+,\p)\G(\nu-iw,\p)\Gamma(\nu-i0^+,\nu-iw;\p)\bigr],
\end{align}
where $f_F(\nu)=1/(e^{\beta\nu}+1)$ is the Fermi-Dirac distribution function.
Now that the resulting expression is regular away from the real axis of $iw\to\omega$, it can be analytically continued into
\begin{align}\label{eq:retarded}
& \chi_\O(\omega+i0^+) = 2\int_{-\infty}^\infty\!\frac{d\nu}{2\pi i}\,f_F(\nu)\int_\p\gamma_\p \notag\\
& \times \bigl[\G_+(\nu+\omega,\p)\G_+(\nu,\p)\Gamma(\nu+\omega+i0^+,\nu+i0^+;\p) \notag\\
&\quad - \G_+(\nu+\omega,\p)\G_-(\nu,\p)\Gamma(\nu+\omega+i0^+,\nu-i0^+;\p) \notag\\
&\quad + \G_+(\nu,\p)\G_-(\nu-\omega,\p)\Gamma(\nu+i0^+,\nu-\omega-i0^+;\p) \notag\\
&\quad - \G_-(\nu,\p)\G_-(\nu-\omega,\p)\Gamma(\nu-i0^+,\nu-\omega-i0^+;\p)\bigr],
\end{align}
where $\G_\pm(\nu,\p)\equiv\G(\nu\pm i0^+,\p)$ are the retarded (upper sign) and advanced (lower sign) fermion propagators.

The retarded correlation function obtained in Eq.~(\ref{eq:retarded}) is naively $O(z)$ because $\G_\pm(\nu,\p)=G(\nu\pm i0^+,\p)+O(z)$, $\Gamma(\nu,\nu';\p)=\gamma_\p+O(z)$, and the chemical potentials are eliminated from the fermion propagators by shifting the integration variable as $\nu\to\nu-\mu$, producing $f_F(\nu-\mu)=ze^{-\beta\nu}+O(z^2)$.
However, such a naive counting breaks down in the zero-frequency limit, $\omega\to0$, where the product of retarded and advanced fermion propagators is decomposed into partial fractions as
\begin{align}
\G_+(\nu,\p)\G_-(\nu,\p) = \frac{\Im[\G(\nu+i0^+,\p)]}{\Im[\Sigma(\nu+i0^+,\p)]}.
\end{align}
Because of $\Sigma(\nu+i0^+,\p)\sim O(z)$ and $\Im[\G(\nu+i0^+,\p)]=-\pi\,\delta(\nu-\ep+\mu)+O(z)$ in the high-temperature limit, we find that
\begin{align}\label{eq:pinch}
\G_+(\nu,\p)\G_-(\nu,\p) = -\frac{\pi\,\delta(\nu-\ep+\mu)}{\Im[\Sigma(\ep-\mu+i0^+,\p)]} + O(z^0)
\end{align}
is inversely proportional to the fugacity so as to promote its order by one.
Therefore, the resulting transport coefficient in Eq.~(\ref{eq:kubo}) actually involves an $O(z^0)$ contribution provided by
\begin{align}\label{eq:transport}
\sigma_\O = 2\beta\int_\p e^{-\beta\ep}\gamma_\p
\frac{z\,\Gamma_{+-}(\p)}{-2\Im[\Sigma_+(\p)]} + O(z).
\end{align}
Here, shorthand notations for on-shell $\Sigma_+(\p)\equiv\Sigma(\ep-\mu+i0^+,\p)$ and $\Gamma_{+-}(\p)\equiv\Gamma(\ep-\mu+i0^+,\ep-\mu-i0^+;\p)$ are introduced.
We note that the latter is real and the former has both real and imaginary parts at $O(z)$ according to Eq.~(\ref{eq:self-energy}).

An important lesson to be learned is as follows:
The correlation function in Fig.~\ref{fig:correlation} is naively $O(z)$ as also seen diagrammatically because there exists one fermion propagator running backward in imaginary time~\cite{Leyronas:2011,Hofmann:2020}.
However, the product of two fermion propagators with the same frequency and wave vector, such as the two thick lines in Fig.~\ref{fig:correlation}, produces an inverse power of the fugacity so as to promote its order by one compared to the naive counting.
This is the so-called pinch singularity~\cite{Eliashberg:1962,Jeon:1995,Jeon:1996,Hidaka:2011}, which needs to be taken into account in order to compute the static transport coefficient correctly to the lowest order in fugacity.

\section{Toward the kinetic theory}\label{sec:kinetics}
\subsection{Vertex function}\label{sec:vertex}
\begin{figure}[t]
\includegraphics[width=0.7\columnwidth]{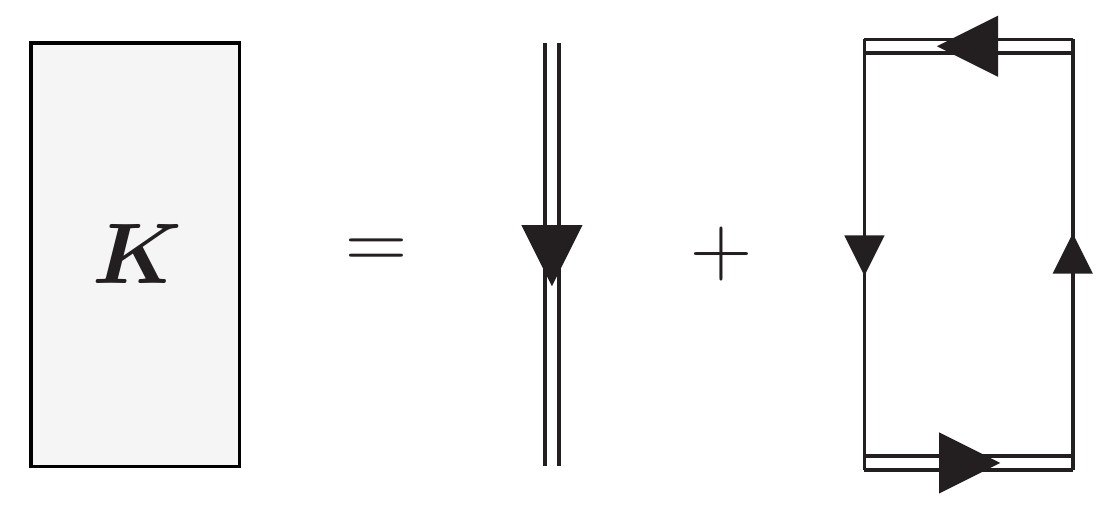}
\caption{\label{fig:kernel}
Maki-Thompson (left) and Aslamazov-Larkin (right) diagrams for the four-point function represented by the rectangle.}
\end{figure}

\begin{figure}[t]
\includegraphics[width=\columnwidth]{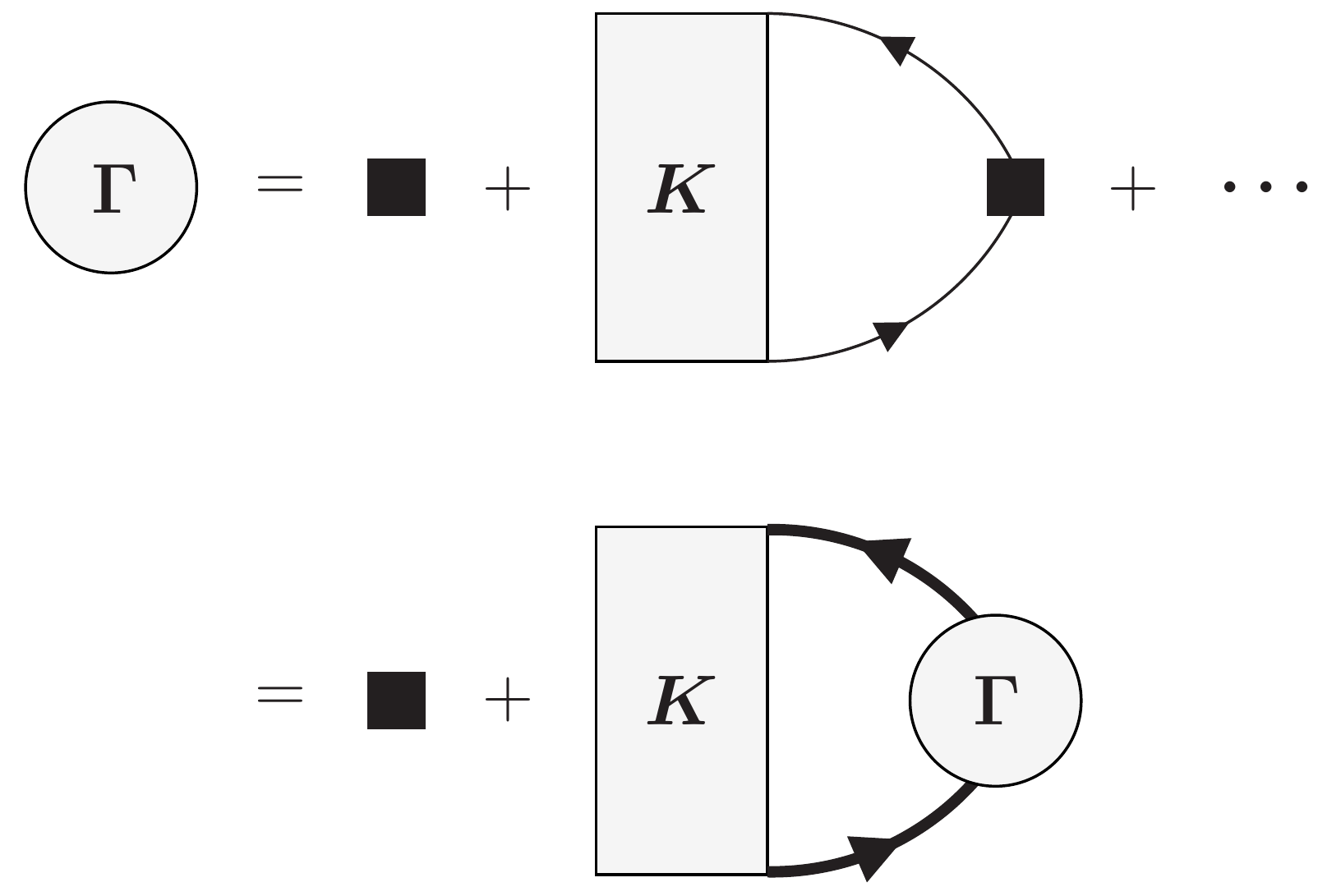}
\caption{\label{fig:vertex}
Vertex function in the naive expansion with respect to the fugacity up to $O(z)$ (top).
An infinite sequence of comparable diagrams needs to be resumed in the zero-frequency limit due to the pinch singularity, which leads to a self-consistent equation for the vertex function (bottom).}
\end{figure}

Our remaining task is to determine the full vertex function to the lowest order in fugacity.
If the naive expansion with respect to the fugacity is applied at nonzero frequency, its leading term is simply the bare vertex function and the next-to-leading-order corrections at $O(z)$ are provided by two types of diagrams called Maki-Thompson and Aslamazov-Larkin~\cite{Enss:2011,Nishida:2019,Hofmann:2020}.%
\footnote{The self-energy diagrams are already taken into account by adopting the full fermion propagator in Eq.~(\ref{eq:time-ordered}).}
The resulting vertex function is expressed as
\begin{align}\label{eq:naive}
& \Gamma(iv+iw,iv;\p) \notag\\
&= \gamma_\p + \frac1\beta\sum_{v'}\int_{\p'}K(iv+iw,iv;\p|iv'+iw,iv';\p') \notag\\
&\quad \times G(iv'+iw,\p')G(iv',\p')\gamma_{\p'} + O(z^2),
\end{align}
where $K(\,*\,)=K_\MT(\,*\,)+K_\AL(\,*\,)$ is the four-point function depicted in Fig.~\ref{fig:kernel} consisting of
\begin{align}\label{eq:kernel_MT}
& K_\MT(iv+iw,iv;\p|iv'+iw,iv';\p') \notag\\
&= D(iv+iv'+iw,\p+\p')
\end{align}
and
\begin{align}\label{eq:kernel_AL}
& K_\AL(iv+iw,iv;\p|iv'+iw,iv';\p') \notag\\
&= -\frac2\beta\sum_{w''}\int_{\p''}D(iw''+iw,\p'')D(iw'',\p'') \notag\\
&\quad \times G(iw''-iv,\p''-\p)G(iw''-iv',\p''-\p').
\end{align}
The diagrammatic representation of such a naive expansion is depicted in Fig.~\ref{fig:vertex} (top), where the second term is seen to be $O(z)$ because each diagram therein can be organized so that there exists one fermion propagator running backward in imaginary time~\cite{Hofmann:2020}.

As learned in Sec.~\ref{sec:pinch}, the above naive expansion breaks down in the zero-frequency limit.
This is because the second term of Eq.~(\ref{eq:naive}) involves the product of two fermion propagators with the same frequency and wave vector.
With the bare fermion propagators therein replaced by the full ones, the pinch singularity produces $z/\Im\Sigma\sim O(z^0)$ comparable to the first term.
Furthermore, an iteration of the same diagrammatic structure for $n$ times also contributes $(z/\Im\Sigma)^n\sim O(z^0)$, so that an infinite sequence of such comparable diagrams needs to be resummed.%
\footnote{Previously, the resummation was performed only approximately assuming a simple geometric series based on the lowest two terms in fugacity~\cite{Enss:2011,Nishida:2019,Hofmann:2020}, which turns out to correspond to the relaxation-time approximation (see Appendix~\ref{app:relaxation}).}
Such a resummation can formally be achieved by replacing the bare vertex function in the second term of Eq.~(\ref{eq:naive}) by the full one, leading to
\begin{align}\label{eq:vertex}
& \Gamma(iv+iw,iv;\p) \notag\\
&= \gamma_\p + \frac1\beta\sum_{v'}\int_{\p'}K(iv+iw,iv;\p|iv'+iw,iv';\p') \notag\\
&\quad \times \G(iv'+iw,\p')\G(iv',\p')\Gamma(iv'+iw,iv';\p').
\end{align}
This is the closed integral equation that self-consistently determines the vertex function to the lowest order in fugacity and its diagrammatic representation is depicted in Fig.~\ref{fig:vertex} (bottom).

\subsection{Analytic continuation}
\subsubsection{Maki-Thompson}
We now evaluate the above self-consistent equation for the vertex function so that it can be analytically continued into that for $\Gamma_{+-}(\p)$ needed to compute the transport coefficient according to Eq.~(\ref{eq:transport}).
Let us start with the Maki-Thompson part denoted by
\begin{align}
& \Gamma_\MT(iv+iw,iv;\p) \notag\\
&\equiv \frac1\beta\sum_{v'}\int_{\p'}K_\MT(iv+iw,iv;\p|iv'+iw,iv';\p') \notag\\
&\quad \times \G(iv'+iw,\p')\G(iv',\p')\Gamma(iv'+iw,iv';\p'),
\end{align}
where the four-point function is provided by Eq.~(\ref{eq:kernel_MT}).

With the Matsubara frequency summation replaced by the complex contour integration over $iv'\to\nu'$, the integrand may have singularities only along $\Im(\nu')=0,-w,-v-w$ in the complex plane of $\nu'$.
Therefore, its contour is deformed into six horizontal lines in a similar way to Fig.~\ref{fig:contour}.
Because only two of them along $\Im(\nu')=-0^+,-w+0^+$ turn out to contribute $O(z^0)$ in the zero-frequency limit, we obtain
\begin{align}
& \Gamma_\MT(iv+iw,iv;\p)
= \int_{-\infty}^\infty\!\frac{d\nu'}{2\pi i}\,f_F(\nu')\int_{\p'} \notag\\
& \times \bigl[D(\nu'+iv+iw,\p+\p')\G(\nu'+iw,\p') \notag\\
&\quad \times \G(\nu'-i0^+,\p')\Gamma(\nu'+iw,\nu'-i0^+;\p') \notag\\
& - D(\nu'+iv,\p+\p')\G(\nu'+i0^+,\p') \notag\\
&\quad \times \G(\nu'-iw,\p')\Gamma(\nu'+i0^+,\nu'-iw;\p')\big] + O(z).
\end{align}
Then, the analytic continuation of $iv\to\ep-\mu-i0^+$ followed by $iw\to i0^+$ leads to
\begin{align}\label{eq:vertex_MT}
& \Gamma_\MT(\ep-\mu+i0^+,\ep-\mu-i0^+;\p) \notag\\
&= -\int_{\q,\p',\q'}e^{-\beta\eq}\,W(\p,\q|\p',\q')
\frac{z\,\Gamma_{+-}(\q)}{-2\Im[\Sigma_+(\q)]} + O(z),
\end{align}
where the pinch singularity in Eq.~(\ref{eq:pinch}) is applied as well as Eq.~(\ref{eq:optical}) with some change of the integration variable.

\subsubsection{Aslamazov-Larkin}
We next turn to the Aslamazov-Larkin part denoted by
\begin{align}
& \Gamma_\AL(iv+iw,iv;\p) \notag\\
&\equiv \frac1\beta\sum_{v'}\int_{\p'}K_\AL(iv+iw,iv;\p|iv'+iw,iv';\p') \notag\\
&\quad \times \G(iv'+iw,\p')\G(iv',\p')\Gamma(iv'+iw,iv';\p').
\end{align}
Here, the four-point function in Eq.~(\ref{eq:kernel_AL}) is evaluated to the lowest order in fugacity as
\begin{align}
& K_\AL(iv+iw,iv;\p|iv'+iw,iv';\p') \notag\\
&= -2z\int_{\p''}e^{-\beta\eps_{\p''-\p}}D(iv+iw+\eps_{\p''-\p}-\mu,\p'') \notag\\
&\quad \times \frac{D(iv+\eps_{\p''-\p}-\mu,\p'')}{iv-iv'+\eps_{\p''-\p}-\eps_{\p''-\p'}} \notag\\
& - 2z\int_{\p''}e^{-\beta\eps_{\p''-\p'}}D(iv'+iw+\eps_{\p''-\p'}-\mu,\p'') \notag\\
&\quad \times \frac{D(iv'+\eps_{\p''-\p'}-\mu,\p'')}{iv'-iv+\eps_{\p''-\p'}-\eps_{\p''-\p}}
+ O(z^2),
\end{align}
which is dominated by the contributions from the poles of the fermion propagators because the branch cuts of the pair propagators contribute $O(z^2)$.

With the Matsubara frequency summation replaced by the complex contour integration over $iv'\to\nu'$, the integrand may have singularities only along $\Im(\nu')=0,-w$ in addition to a pole at $\nu'=iv+\eps_{\p''-\p}-\eps_{\p''-\p'}$ in the complex plane of $\nu'$.
Therefore, its contour is deformed into four horizontal lines as in Fig.~\ref{fig:contour} and one clockwise circle around the pole.
Because only the latter turns out to contribute $O(z^0)$ in the zero-frequency limit, we obtain
\begin{align}
& \Gamma_\AL(iv+iw,iv;\p) = 2z\int_{\p',\p''}e^{-\beta\eps_{\p''-\p}} \notag\\
& \times D(iv+iw+\eps_{\p''-\p}-\mu,\p'')D(iv+\eps_{\p''-\p}-\mu,\p'') \notag\\
& \times \G(\nu'+iw,\p')\G(\nu',\p') \notag\\
& \times \Gamma(\nu'+iw,\nu';\p')|_{\nu'\to iv+\eps_{\p''-\p}-\eps_{\p''-\p'}} + O(z).
\end{align}
Then, the analytic continuation of $iv\to\ep-\mu-i0^+$ followed by $iw\to i0^+$ leads to
\begin{align}\label{eq:vertex_AL}
& \Gamma_\AL(\ep-\mu+i0^+,\ep-\mu-i0^+;\p) \notag\\
&= 2\int_{\q,\p',\q'}e^{-\beta\eq}\,W(\p,\q|\p',\q')
\frac{z\,\Gamma_{+-}(\p')}{-2\Im[\Sigma_+(\p')]} + O(z),
\end{align}
where the pinch singularity in Eq.~(\ref{eq:pinch}) is applied as well as Eq.~(\ref{eq:transition}) with some change of the integration variable.

\subsection{On-shell self-consistent equation}
With Eqs.~(\ref{eq:vertex_MT}) and (\ref{eq:vertex_AL}) substituted, Eq.~(\ref{eq:vertex}) is analytically continued into
\begin{align}\label{eq:vertex_on-shell}
& \Gamma_{+-}(\p) = \gamma_\p
- \int_{\q,\p',\q'}e^{-\beta\eq}\,W(\p,\q|\p',\q') \notag\\
&\quad \times \left\{\frac{z\,\Gamma_{+-}(\q)}{-2\Im[\Sigma_+(\q)]}
- 2\frac{z\,\Gamma_{+-}(\p')}{-2\Im[\Sigma_+(\p')]}\right\} + O(z),
\end{align}
which is now the closed integral equation for the on-shell vertex function to the lowest order in fugacity.

In order to show that the resulting on-shell self-consistent equation is identical to the linearized Boltzmann equation~\cite{Jeon:1995,Jeon:1996,Hidaka:2011}, we introduce
\begin{align}
\varphi(\p) \equiv \frac{z\,\Gamma_{+-}(\p)}{-2\Im[\Sigma_+(\p)]},
\end{align}
so that Eq.~(\ref{eq:vertex_on-shell}) turns into
\begin{align}
\gamma_\p &= \frac{-2\Im[\Sigma_+(\p)]}{z}\varphi(\p)
+ \int_{\q,\p',\q'}e^{-\beta\eq}\,W(\p,\q|\p',\q') \notag\\
&\quad \times [\varphi(\q)-2\varphi(\p')] + O(z).
\end{align}
Because of $W(\p,\q|\p',\q')=W(\p,\q|\q',\p')$ and
\begin{align}
-2\Im[\Sigma_+(\p)] = z\int_{\q,\p',\q'}e^{-\beta\eq}\,W(\p,\q|\p',\q') + O(z^2)
\end{align}
according to Eqs.~(\ref{eq:transition}), (\ref{eq:optical}), and (\ref{eq:self-energy}), we indeed find that
\begin{align}\label{eq:on-shell}
\gamma_\p &= \int_{\q,\p',\q'}e^{-\beta\eq}\,W(\p,\q|\p',\q') \notag\\
&\quad \times [\varphi(\p)+\varphi(\q)-\varphi(\p')-\varphi(\q')] + O(z)
\end{align}
is none other than the linearized Boltzmann equation~\cite{Lifshitz-Pitaevskii} (see also Appendix~\ref{app:boltzmann}).

Once the solution of $\varphi(\p)$ for a given $\gamma_\p$ is determined, the corresponding transport coefficient in Eq.~(\ref{eq:transport}) is provided by
\begin{align}\label{eq:coefficient}
\sigma_\O = 2\beta\int_\p e^{-\beta\ep}\gamma_\p\,\varphi(\p) + O(z).
\end{align}
Therefore, it is hereby established that computing the shear viscosity and the thermal conductivity in the high-temperature limit to the lowest order in fugacity is reduced to the kinetic theory, which contrasts with the bulk viscosity and constitutes the main outcome of this paper.

\section{Transport coefficients}\label{sec:transport}
Finally, we solve the linearized Boltzmann equation in Eq.~(\ref{eq:on-shell}) to compute the transport coefficients according to Eq.~(\ref{eq:coefficient}) both in two and three dimensions.%
\footnote{There exists no solution to Eq.~(\ref{eq:on-shell}) for $d=1$ because its right-hand side vanishes identically due to the energy and momentum conservations.}
To this end, we expand $\varphi(\p)$ in terms of the generalized Laguerre polynomials,
\begin{align}
\varphi(\p) = \beta\,\frac{p_xp_y}{m}\sum_{n=0}^{N-1}c_nL_n^{d/2+1}(\beta\ep)
\end{align}
for the shear viscosity with $\gamma_\p=p_xp_y/m$ and
\begin{align}\label{eq:expansion}
\varphi(\p) = \frac{p_x}{m}\sum_{n=1}^Nc_nL_n^{d/2}(\beta\ep)
\end{align}
for the thermal conductivity with $\gamma_\p=[\ep-(\E+\P)/\N]\,p_x/m$~\cite{Lifshitz-Pitaevskii}.%
\footnote{The $n=0$ term in Eq.~(\ref{eq:expansion}) is not needed because it vanishes when substituted into Eq.~(\ref{eq:on-shell}) due to the momentum conservation.
This is also consistent with the Chapman-Enskog condition in Eq.~(\ref{eq:condition}) from the kinetic theory perspective.}
We note that $\N=2z/\lambda_T^d+O(z^2)$, $\E=(d/2)\,\N/\beta+O(z^2)$, and $\P=\N/\beta+O(z^2)$ in the high-temperature limit, where $\lambda_T=\sqrt{2\pi\beta/m}$ is the thermal de Broglie wavelength~\cite{Nishida:2019}.
If the above expansion is truncated up to the lowest $N$ terms, the simplest case of $N=1$ corresponds to the relaxation-time approximation to be described in Appendix~\ref{app:relaxation}.

\begin{figure}[t]
\includegraphics[width=\columnwidth]{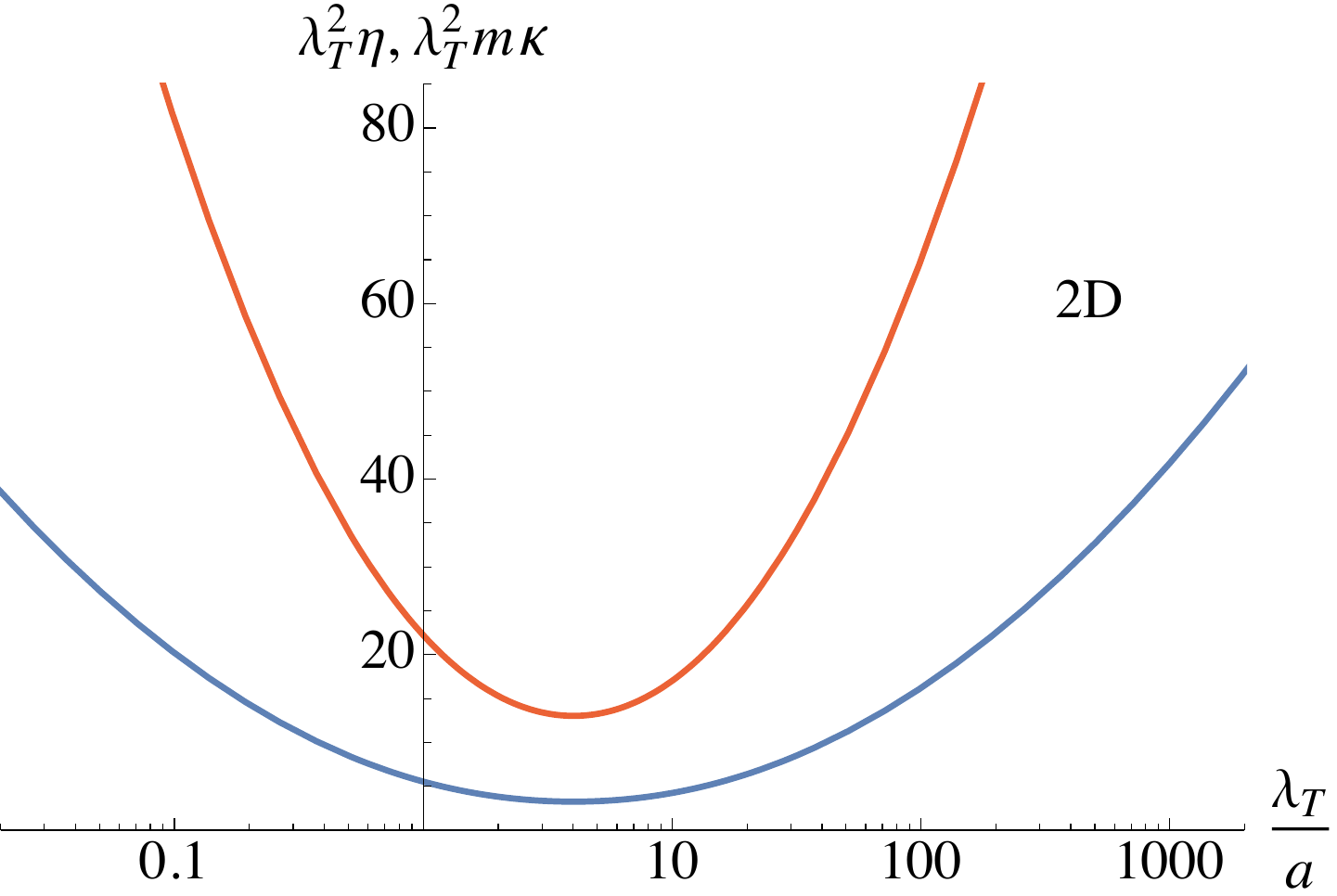}\bigskip\\
\includegraphics[width=\columnwidth]{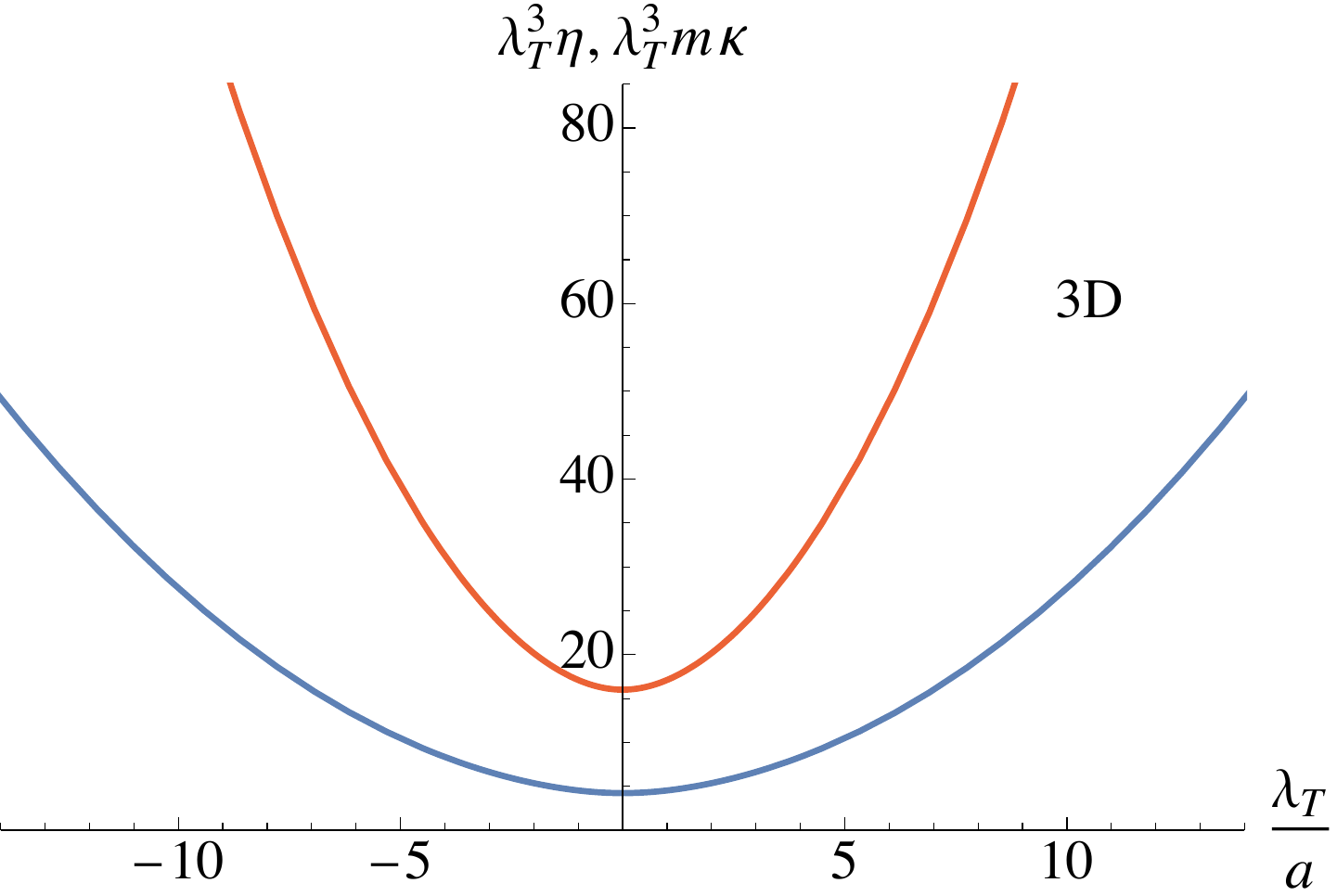}
\caption{\label{fig:shear-thermal}
Shear viscosity and thermal conductivity in the high-temperature limit for $d=2$ (top) and $d=3$ (bottom) as functions of $\lambda_T/a$ in the forms of $\lambda_T^d\eta$ (blue lower curves) and $\lambda_T^dm\kappa$ (red upper curves).
Both of them are even functions for $d=3$.}
\end{figure}

Here, $N$ is increased up to $N=10$, which is confirmed to be more than sufficient for convergence of the presented results, and the resulting shear viscosity and thermal conductivity for $d=2$ and $d=3$ are plotted in Fig.~\ref{fig:shear-thermal} as functions of the inverse scattering length.
They are exact in the high-temperature limit to the lowest order in fugacity and found to be slightly larger than those obtained with the relaxation-time approximation up to 2.5\% deviations.
In particular, we find that the shear viscosity and the thermal conductivity for $d=2$ reach their minima of $\lambda_T^d\eta=3.257$ at $\lambda_T/a=4.019$ and $\lambda_T^dm\kappa=13.03$ at $\lambda_T/a=4.017$, respectively, and our thermal conductivity in two dimensions is new to the best of our knowledge.
We also find $\lambda_T^d\eta=4.231$ and $\lambda_T^dm\kappa=16.01$ for $d=3$ at infinite scattering length, $\lambda_T/a=0$, where the relaxation-time approximation produces $\lambda_T^d\eta=15\pi/(8\sqrt2)\approx4.165$~\cite{Massignan:2005,Bruun:2005} and $\lambda_T^dm\kappa=225\pi/(32\sqrt2)\approx15.62$~\cite{Braby:2010}, respectively.

\begin{figure}[t]
\includegraphics[width=\columnwidth]{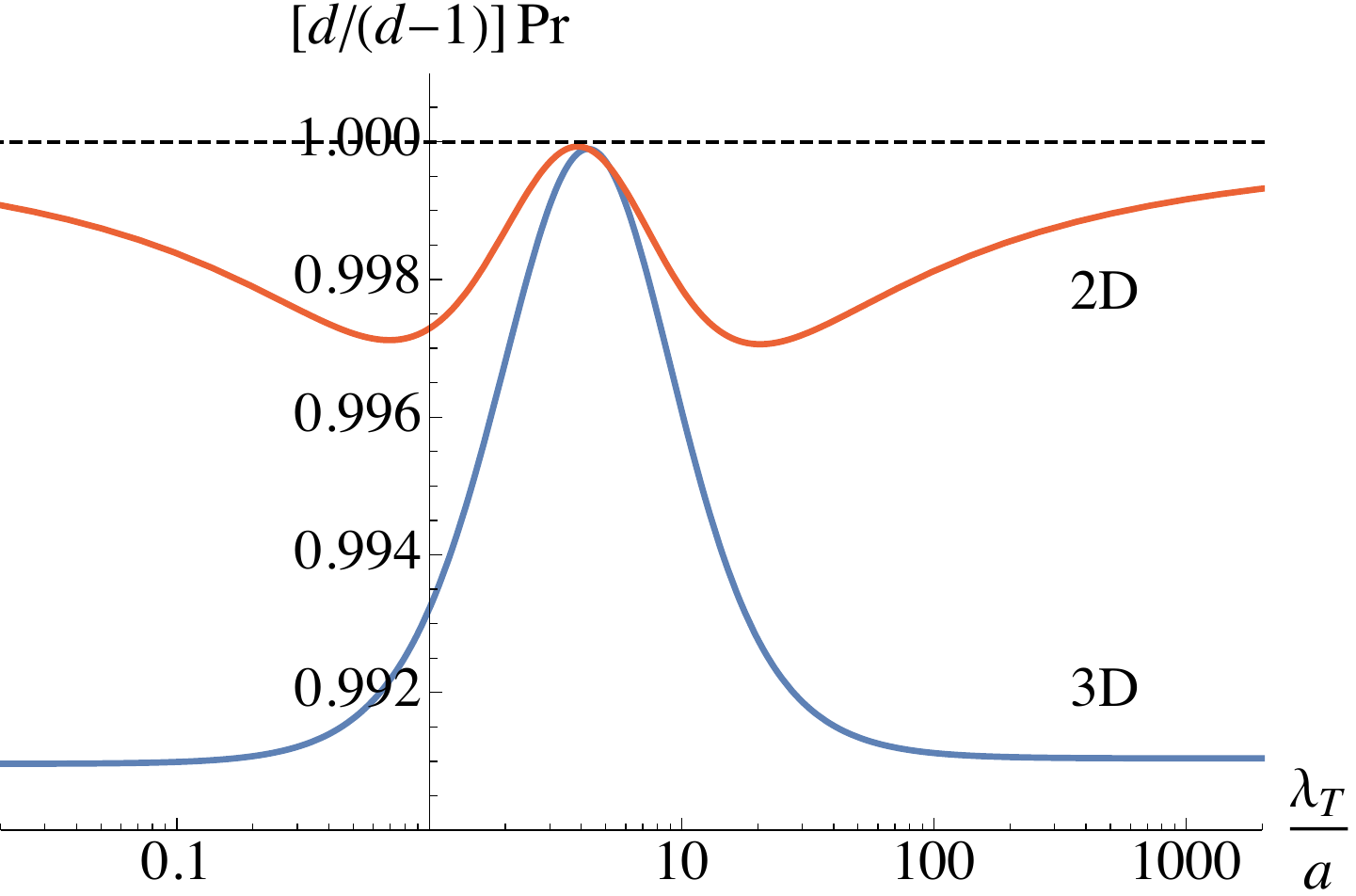}
\caption{\label{fig:prandtl}
Prandtl number in the high-temperature limit for $d=2$ (red upper curve) and $d=3$ (blue lower curve) as a function of $\lambda_T/a$ in the form of $[d/(d-1)]\,\Pr$.
The horizontal dashed line indicates the constant value in the relaxation-time approximation in Eq.~(\ref{eq:constant}).}
\end{figure}

Our exact results for the shear viscosity and the thermal conductivity allow us to study the Prandtl number defined by their ratio via
\begin{align}\label{eq:prandtl}
\Pr \equiv \frac\eta{m\kappa}c_\P.
\end{align}
Here, the heat capacity at constant pressure per particle is provided by $c_\P=(d+2)/2+O(z)$ in the high-temperature limit~\cite{Braby:2010}.
The resulting Prandtl number for $d=2$ and $d=3$ is plotted in Fig.~\ref{fig:prandtl} as a function of the inverse scattering length.
Remarkably, it is found to exhibit the nonmonotonic behavior slightly below the constant value of $\Pr=(d-1)/d$ in the relaxation-time approximation.
In particular, it reaches its two minima of $\Pr=0.4986$ and $0.4985$ at $\lambda_T/a=0.6948$ and $20.45$, respectively, for $d=2$ and its maximum of $\Pr=0.5000$ at $\lambda_T/a=3.871$ for $d=2$ and $\Pr=0.6666$ at $\lambda_T/a=\pm4.262$ for $d=3$.
We also find that the minimum of $\Pr=0.6606$ is reached at infinite scattering length, $\lambda_T/a=0$, for $d=3$.

\section{Summary}\label{sec:summary}
In this paper, we evaluated the Kubo formula [Eq.~(\ref{eq:kubo})] for the shear viscosity and the thermal conductivity exactly in the high-temperature limit to the lowest order in fugacity and showed that it is reduced to the linearized Boltzmann equation [Eqs.~(\ref{eq:on-shell}) and (\ref{eq:coefficient})].
This task was achieved by resuming all contributions that are naively higher orders in fugacity but become comparable in the zero-frequency limit due to the pinch singularity [Eq.~(\ref{eq:pinch})].
Consequently, the complete correspondence between the microscopic and kinetic theories for the transport coefficients is now established beyond the previous relaxation-time approximation.
The bulk viscosity is however the exception because its corresponding operator in the Kubo formula is essentially a two-body operator without the one-body part in the form of Eq.~(\ref{eq:operator}).
We also found by solving the linearized Boltzmann equation numerically for an arbitrary scattering length that the Prandtl number in the high-temperature limit exhibits the nonmonotonic behavior slightly below the constant value in the relaxation-time approximation both in two and three dimensions (Fig.~\ref{fig:prandtl}).

Here, we worked with the Matsubara formalism to take advantage of the established quantum virial expansion, which however requires the cumbersome analytic continuation~\cite{Eliashberg:1962}.
Instead, it will also be worthwhile to adopt the Keldysh formalism, where it is possible to directly identify diagrams with the pinch singularity and formulate the systematic expansion incorporating higher-order corrections~\cite{Hidaka:2011}.
Developing the quantum virial expansion in the Keldysh formalism may provide us with a useful theoretical framework to study broad nonequilibrium phenomena in ultracold-atom physics.

\acknowledgments
The authors thank Yoshimasa Hidaka for giving an informal lecture on his work~\cite{Hidaka:2011}.
This work was supported by JSPS KAKENHI Grants No.\ JP19J13698 and No.\ JP18H05405.

\appendix
\section{Spectral representation of three-point functions}\label{app:spectral}
Let us review the spectral representation of three-point functions in the form of
\begin{align}
J(\tau_1,\tau_2,\tau_3) = \<\T\,\psi^\+(\tau_1)\psi(\tau_2)\O(\tau_3)\>,
\end{align}
which is relevant to the full vertex function introduced in Eq.~(\ref{eq:time-ordered}).
Here, its spin indices and spatial coordinates are suppressed for the sake of brevity and this section is partly based on Ref.~\cite{Eliashberg:1962} (see Appendix therein).

The above imaginary-time-ordered product consists of six terms corresponding to different orderings of three imaginary times and it will turn out to be convenient to group them into two cycles as
\begin{align}
J(\tau_1,\tau_2,\tau_3) = J_{123}(\tau_1,\tau_2,\tau_3) + J_{213}(\tau_1,\tau_2,\tau_3),
\end{align}
where
\begin{align}
J_{123}(\tau_1,\tau_2,\tau_3)
&\equiv J(\tau_1,\tau_2,\tau_3)[\Theta(\tau_1,\tau_2,\tau_3) \notag\\
&\quad + \Theta(\tau_2,\tau_3,\tau_1) + \Theta(\tau_3,\tau_1,\tau_2)]
\end{align}
and
\begin{align}
J_{213}(\tau_1,\tau_2,\tau_3)
&\equiv J(\tau_1,\tau_2,\tau_3)[\Theta(\tau_2,\tau_1,\tau_3) \notag\\
&\quad + \Theta(\tau_1,\tau_3,\tau_2) + \Theta(\tau_3,\tau_2,\tau_1)]
\end{align}
with $\Theta(\tau_1,\tau_2,\tau_3)\equiv\theta(\tau_1-\tau_2)\theta(\tau_2-\tau_3)$ being a step function of three variables.
By inserting identity operators composed of the Hamiltonian eigenstates, $J_{123}(\tau_1,\tau_2,\tau_3)$ can be expressed as
\begin{align}
& J_{123}(\tau_1,\tau_2,\tau_3) = \frac1Z\sum_{l,m,n}\J_{lmn} \notag\\
& \times e^{-E_l(\tau_3-\tau_1)-E_m(\tau_1-\tau_2)-E_n(\tau_2-\tau_3)}
\bigl[e^{-\beta E_l}\Theta(\tau_1,\tau_2,\tau_3) \notag\\
&\quad - e^{-\beta E_m}\Theta(\tau_2,\tau_3,\tau_1)
+ e^{-\beta E_n}\Theta(\tau_3,\tau_1,\tau_2)\bigr],
\end{align}
where $\J_{lmn}\equiv\<l|\psi^\+|m\>\<m|\psi|n\>\<n|\O|l\>$ is a product of three matrix elements.
A similar expression is obtained for $J_{213}(\tau_1,\tau_2,\tau_3)$ by exchanging $\psi^\+\leftrightarrow\psi$ in $-J_{123}(\tau_2,\tau_1,\tau_3)$.

Then, we would like to evaluate its Fourier component provided by
\begin{align}
J_{123}(iv,iv',iw) &= \int_0^\beta\!d\tau_3\int_0^\beta\!d\tau_2\int_0^\beta\!d\tau_1\,
e^{-iv\tau_1+iv'\tau_2-iw\tau_3} \notag\\
&\quad \times J_{123}(\tau_1,\tau_2,\tau_3).
\end{align}
This computation is facilitated by changing the integration variables to $\tau_{12}\equiv\tau_1-\tau_2$, $\tau_{23}\equiv\tau_2-\tau_3$, and $\tau_3$, so that the intervals of integration turn into $\int_0^\beta\!d\tau_3\int_{-\tau_3}^{\beta-\tau_3}\!d\tau_{23}\int_{-\tau_{23}-\tau_3}^{\beta-\tau_{23}-\tau_3}\!d\tau_{12}$.
Because of the periodicities of the integrand with respect to $\tau_{12}$ and $\tau_{23}$ by $\beta$, the integration can be performed instead on the intervals of $\int_0^\beta\!d\tau_3\int_0^\beta\!d\tau_{23}\int_0^\beta\!d\tau_{12}$ assuming the vanishing integrand for $\tau_{12}+\tau_{23}>\beta$.
Therefore, we find
\begin{align}
& J_{123}(iv,iv',iw) \notag\\
&= \frac1Z\sum_{l,m,n}\J_{lmn}
\int_0^\beta\!d\tau_3\int_0^\beta\!d\tau_{23}\int_0^{\beta-\tau_{23}}\!d\tau_{12} \notag\\
&\quad \times e^{-iv(\tau_{12}+\tau_{23}+\tau_3)+iv'(\tau_{23}+\tau_3)-iw\tau_3} \notag\\
&\quad \times e^{(E_l-E_m)\tau_{12}+(E_l-E_n)\tau_{23}-\beta E_l},
\end{align}
which readily leads to
\begin{align}
& J_{123}(iv,iv',iw) = \frac1Z\sum_{l,m,n}\J_{lmn}\,\beta\,\delta_{v',v+w} \notag\\
& \times \biggl[-\frac{e^{-\beta E_l}}{(iw-E_n+E_l)(iv-E_l+E_m)} \notag\\
&\quad + \frac{e^{-\beta E_m}}{(iv-E_l+E_m)(-iv'-E_m+E_n)} \notag\\
&\quad - \frac{e^{-\beta E_n}}{(-iv'-E_m+E_n)(iw-E_n+E_l)}\biggr].
\end{align}
A similar expression is obtained for the Fourier component of $J_{213}(\tau_1,\tau_2,\tau_3)$ by exchanging $\psi^\+\leftrightarrow\psi$ in $-J_{123}(-iv',-iv,iw)$, so that the Fourier component of $J(\tau_1,\tau_2,\tau_3)$ reads
\begin{align}
J(iv,iv',iw) &= J_{123}(iv,iv',iw) \notag\\
&\quad - J_{123}(-iv',-iv,iw)|_{\psi^\+\leftrightarrow\psi}.
\end{align}

Because $J(iv,iv',iw)$ is found to be proportional to $\delta_{v',v+w}$, a function of two independent variables can be introduced via $J(iv,iv',iw)=\beta\,\delta_{v',v+w}\,\tilde{J}(iv,iv+iw)$.
It is now obvious from the above spectral representation that $\tilde{J}(\nu,\nu+iw)$ may have singularities only along $\Im(\nu)=0,-w$ in the complex plane of $\nu$.
In particular, the same is true for the full vertex function, $\Gamma(\nu+iw,\nu;\p)$, because $\tilde{J}(iv,iv+iw)$ is reduced to $\G(iv+iw,\p)\G(iv,\p)\Gamma(iv+iw,iv;\p)$ in Eq.~(\ref{eq:time-ordered}) as a special case.

\section{Boltzmann equation for transport coefficients}\label{app:boltzmann}
Let us review computations of the transport coefficients with the classical Boltzmann equation~\cite{Lifshitz-Pitaevskii}:
\begin{align}\label{eq:boltzmann}
\frac{\d f_\p}{\d t} + \frac{\d\ep}{\d\p}\cdot\frac{\d f_\p}{\d\r}
= \left(\frac{\d f_\p}{\d t}\right)_\coll.
\end{align}
Here, $f_\p=f_\p(t,\r)$ is a local distribution function per spin and the collision term is provided by
\begin{align}
\left(\frac{\d f_\p}{\d t}\right)_\coll
= \int_{\q,\p',\q'}W(\p,\q|\p',\q')(f_{\p'}f_{\q'}-f_\p f_\q)
\end{align}
with the transition rate satisfying $W(\p,\q|\p',\q')=W(\q,\p|\q',\p')=W(\p',\q'|\p,\q)=W(\p+m\v,\q+m\v|\p'+m\v,\q'+m\v)$ and being proportional to $(2\pi)^{d+1}\delta(\ep+\eq-\eps_{\p'}-\eps_{\q'})\delta(\p+\q-\p'-\q')$.

\subsection{Linearization}
When the system is slightly out of thermodynamic equilibrium, the distribution function is expanded as $f_\p=\bar{f}_\p+\delta f_\p$, where
\begin{align}
\bar{f}_\p = \exp\!\left\{-\beta\left[\frac{(\p-m\v)^2}{2m}-\mu\right]\right\}
\end{align}
is a local equilibrium distribution function.
Because the local chemical potential $\mu=\mu(t,\r)$, velocity $\v=\v(t,\r)$, and inverse temperature $\beta=\beta(t,\r)$ are determined so that $\bar{f}_\p$ solely produces local number, momentum, and energy densities, $\delta f_\p$ satisfies the Chapman-Enskog condition of
\begin{align}\label{eq:condition}
\int_\p\delta f_\p = \int_\p p_i\delta f_\p = \int_\p\ep\delta f_\p = 0.
\end{align}
In what follows, implicit summations over repeated spatial indices $i=1,2,\dots,d$ are assumed.

The substitution of $f_\p=\bar{f}_\p+\delta f_\p$ into the left-hand side (LHS) of Eq.~(\ref{eq:boltzmann}) with the help of thermodynamic relations and continuity equations leads to
\begin{align}\label{eq:LHS}
(\LHS) &= \beta\bar{f}_\p\biggl[\pi_{ij}(\p-m\v)\frac{V_{ij}}{2} + \pi(\p-m\v)V \notag\\
&\quad - \bm\Q(\p-m\v)\cdot\grad\ln\beta\biggr] + O(\delta f).
\end{align}
Here, the shear and bulk strain rates are denoted by
\begin{align}
V_{ij} &= \d_iv_j + \d_jv_i - \frac2d\delta_{ij}\grad\cdot\v
\end{align}
and $V=\grad\cdot\v$, respectively, whereas
\begin{align}
\pi_{ij}(\p) = \frac{p_ip_j}{m} - \frac{\p^2}{dm}\delta_{ij}
\end{align}
is the traceless part of the stress tensor,
\begin{align}
\pi(\p) = \frac{\p^2}{dm}
- \left(\frac{\d\P}{\d\N}\right)_\E - \ep\left(\frac{\d\P}{\d\E}\right)_\N
\end{align}
its modified diagonal part, and
\begin{align}
\bm\Q(\p) = \left(\ep-\frac{\E+\P}{\N}\right)\frac\p{m}
\end{align}
the heat flux of a single particle.
We note that $\pi(\p)=0$ because of $\P=(2/d)\,\E=\N/\beta$, indicating the vanishing bulk viscosity within the Boltzmann equation~\cite{Dusling:2013,Chafin:2013}.

On the other hand, the substitution of $f_\p=\bar{f}_\p+\delta f_\p$ into the right-hand side (RHS) of Eq.~(\ref{eq:boltzmann}) with $\delta f_\p\equiv\beta\bar{f}_\p\phi_\p$ introduced leads to
\begin{align}
(\RHS) &= \beta\bar{f}_\p\int_{\q,\p',\q'}\bar{f}_\q\,W(\p,\q|\p',\q') \notag\\
&\quad \times (\phi_{\p'}+\phi_{\q'}-\phi_\p-\phi_\q) + O(\delta f^2)
\end{align}
because the equilibrium distribution function cancels the collision term.
In order for the resulting expression to match the left-hand side in Eq.~(\ref{eq:LHS}) for arbitrary $\mu$, $\v$, and $\beta$, $\phi_\p$ must be in the form of
\begin{align}\label{eq:deviation}
\phi_\p = -e^{-\beta\mu}\left[\varphi_{ij}(\p-m\v)\frac{V_{ij}}{2}
- \bm\varphi(\p-m\v)\cdot\grad\ln\beta\right].
\end{align}
Here, $\varphi_{ij}(\p)$ and $\bm\varphi(\p)$ determine the deviations from the equilibrium distribution function induced by the shear strain rate and the temperature gradient, respectively, which solve
\begin{subequations}\label{eq:linearized}
\begin{align}
&\pi_{xy}(\p) = \int_{\q,\p',\q'}e^{-\beta\eq}\,W(\p,\q|\p',\q') \notag\\
&\quad \times [\varphi_{xy}(\p)+\varphi_{xy}(\q)-\varphi_{xy}(\p')-\varphi_{xy}(\q')]
\end{align}
and
\begin{align}
\Q_x(\p) &= \int_{\q,\p',\q'}e^{-\beta\eq}\,W(\p,\q|\p',\q') \notag\\
&\quad \times [\varphi_x(\p)+\varphi_x(\q)-\varphi_x(\p')-\varphi_x(\q')].
\end{align}
\end{subequations}
This is the linearized Boltzmann equation identical to Eq.~(\ref{eq:on-shell}) derived microscopically from the Kubo formula.

\subsection{Transport coefficients}
The stress tensor and the energy flux in the kinetic theory are provided by
\begin{align}
\Pi_{ij} &= 2\int_\p\frac{p_ip_j}{m}f_\p, \\
\bm\K &= 2\int_\p\ep\frac{\p}{m}f_\p,
\end{align}
respectively, where the spin degeneracy accounts for the prefactor of $2$.
The substitution of $f_\p=\bar{f}_\p+\delta f_\p$ with Eq.~(\ref{eq:condition}) imposed decomposes the former into $\Pi_{ij}=\P\delta_{ij}+m\N v_iv_j-\sigma_{ij}$ with the dissipative correction of
\begin{align}
\sigma_{ij} = -2\beta\int_\p\bar{f}_\p\,\pi_{ij}(\p-m\v)\,\phi_\p,
\end{align}
whereas the latter is decomposed into $\K_i=(\E+\eps_{m\v}\N+\P)v_i-\sigma_{ij}v_j+k_i$ with
\begin{align}
\k = 2\beta\int_\p\bar{f}_\p\,\bm\Q(\p-m\v)\,\phi_\p.
\end{align}
By substituting Eq.~(\ref{eq:deviation}) and comparing the resulting expressions to the corresponding ones in hydrodynamics,
\begin{align}
\sigma_{ij} &= \eta\,V_{ij} + \zeta\,\delta_{ij}V, \\
\k &= -\kappa\,\grad T,
\end{align}
we find $\zeta=0$ for the bulk viscosity and
\begin{subequations}\label{eq:formula}
\begin{align}
\eta &= 2\beta\int_\p e^{-\beta\ep}\pi_{xy}(\p)\varphi_{xy}(\p), \\
T\kappa &= 2\beta\int_\p e^{-\beta\ep}\Q_x(\p)\varphi_x(\p),
\end{align}
\end{subequations}
which are identical to Eq.~(\ref{eq:coefficient}) for the shear viscosity and the thermal conductivity.

\subsection{Relaxation-time approximation}\label{app:relaxation}
Whereas the linearized Boltzmann equation was solved numerically in Sec.~\ref{sec:transport}, it is also common to solve it with the so-called relaxation-time approximation~\cite{Massignan:2005,Bruun:2005,Braby:2010,Bruun:2012,Schafer:2012}, which assumes simple proportional relations in%
\footnote{This is the approximation identifying the collision term with $(\d f_\p/\d t)_\coll=-\delta f_\p/\tau_{\eta,\kappa}$ for each transport coefficient.}
\begin{subequations}\label{eq:ansatz}
\begin{align}
\varphi_{xy}(\p) &= z\tau_\eta\pi_{xy}(\p), \\
\varphi_x(\p) &= z\tau_\kappa\Q_x(\p).
\end{align}
\end{subequations}
Here, $\tau_\eta$ and $\tau_\kappa$ are the shear and thermal relaxation times, respectively, which are determined by substituting the above relations into the linearized Boltzmann equation in Eqs.~(\ref{eq:linearized}).
For our system with the transition rate provided by Eq.~(\ref{eq:transition}), they are found to be
\begin{align}\label{eq:relaxation}
\frac\beta{z\tau_\eta}
&= \frac\pi{(2\sqrt2\,\pi)^d\Gamma(d/2)\Gamma(d/2+2)}
\int_0^\infty\!d\eps\,\eps^de^{-\eps} \notag\\
&\quad \times \left|\frac{(d-2)\Omega_{d-1}}
{(\sqrt{2\pi}\,a/\lambda_T)^{2-d}-(-\eps-i0^+)^{d/2-1}}\right|^2
\end{align}
and $\tau_\kappa=[d/(d-1)]\,\tau_\eta$.

Then, the substitution of Eqs.~(\ref{eq:ansatz}) into Eqs.~(\ref{eq:formula}) relates the shear viscosity and the thermal conductivity to their relaxation times as
\begin{align}
\eta = \frac{2z\tau_\eta}{\beta\lambda_T^d}, \qquad
\kappa = \frac{(d+2)z\tau_\kappa}{m\beta\lambda_T^d}.
\end{align}
Because the former with Eq.~(\ref{eq:relaxation}) agrees with the shear viscosity obtained in Refs.~\cite{Enss:2011,Nishida:2019,Hofmann:2020}, the approximate resummation scheme adopted therein assuming a simple geometric series based on the lowest two terms in fugacity turns out to correspond to the relaxation-time approximation.
We also confirm that the Prandtl number defined in Eq.~(\ref{eq:prandtl}) reads
\begin{align}\label{eq:constant}
\Pr = \frac{d-1}{d},
\end{align}
taking the constant value independent of the scattering length~\cite{Braby:2010,Frank:2020}, which is however true only in the relaxation-time approximation.


\begin{thebibliography}{99}

\bibitem{Bloch:2008}
I.~Bloch, J.~Dalibard, and W.~Zwerger,
``Many-body physics with ultracold gases,''
\href{https://doi.org/10.1103/RevModPhys.80.885}
{Rev.\ Mod.\ Phys.\ \textbf{80}, 885-964 (2008)}.
%arXiv:0704.3011 [cond-mat.other]

\bibitem{Giorgini:2008}
S.~Giorgini, L.~P.~Pitaevskii, and S.~Stringari,
``Theory of ultracold atomic Fermi gases,''
\href{https://doi.org/10.1103/RevModPhys.80.1215}
{Rev.\ Mod.\ Phys.\ \textbf{80}, 1215-1274 (2008)}.
%arXiv:0706.3360 [cond-mat.other]

\bibitem{Zwerger:2012}
\textit{The BCS-BEC Crossover and the Unitary Fermi Gas,}
edited by W.~Zwerger,
\href{https://doi.org/10.1007/978-3-642-21978-8}
{Lecture Notes in Physics (Springer, Berlin, 2012), Vol.\ 836}.

\bibitem{Randeria:2014}
M.~Randeria and E.~Taylor,
``Crossover from Bardeen-Cooper-Schrieffer to Bose-Einstein condensation and the unitary Fermi gas,''
\href{https://doi.org/10.1146/annurev-conmatphys-031113-133829}
{Annu.\ Rev.\ Condens.\ Matter Phys.\ \textbf{5}, 209-232 (2014)}.
%arXiv:1306.5785 [cond-mat.quant-gas]

\bibitem{Mehen:2000}
T.~Mehen, I.~W.~Stewart, and M.~B.~Wise,
``Conformal invariance for non-relativistic field theory,''
\href{https://doi.org/10.1016/S0370-2693(00)00006-X}
{Phys.\ Lett.\ B \textbf{474}, 145-152 (2000)}.
%arXiv:hep-th/9910025

\bibitem{Son:2006}
D.~T.~Son and M.~Wingate,
``General coordinate invariance and conformal invariance in nonrelativistic physics: Unitary Fermi gas,''
\href{https://doi.org/10.1016/j.aop.2005.11.001}
{Ann.\ Phys.\ (NY) \textbf{321}, 197-224 (2006)}.
%arXiv:cond-mat/0509786 [cond-mat.other]

\bibitem{Nishida:2007}
Y.~Nishida and D.~T.~Son,
``Nonrelativistic conformal field theories,''
\href{https://doi.org/10.1103/PhysRevD.76.086004}
{Phys.\ Rev.\ D \textbf{76}, 086004 (2007)}.
%arXiv:0706.3746 [hep-th]

\bibitem{Cao:2011a}
C.~Cao, E.~Elliott, J.~Joseph, H.~Wu, J.~Petricka, T.~Sch\"afer, and J.~E.~Thomas,
``Universal quantum viscosity in a unitary Fermi gas,''
\href{https://doi.org/10.1126/science.1195219}
{Science \textbf{331},58-61 (2011)}.
%arXiv:1007.2625 [cond-mat.quant-gas]

\bibitem{Cao:2011b}
C.~Cao, E.~Elliott, H.~Wu, and J.~E.~Thomas,
``Searching for perfect fluids: Quantum viscosity in a universal Fermi gas,''
\href{https://doi.org/10.1088/1367-2630/13/7/075007}
{New J.\ Phys.\ \textbf{13}, 075007 (2011)}.
%arXiv:1105.2496 [cond-mat.quant-gas]

\bibitem{Elliott:2014b}
E.~Elliott, J.~A.~Joseph, and J.~E.~Thomas,
``Anomalous minimum in the shear viscosity of a Fermi gas,''
\href{https://doi.org/10.1103/PhysRevLett.113.020406}
{Phys.\ Rev.\ Lett.\ \textbf{113}, 020406 (2014)}.
%arXiv:1311.2049 [cond-mat.quant-gas]

\bibitem{Joseph:2015}
J.~A.~Joseph, E.~Elliott, and J.~E.~Thomas,
``Shear viscosity of a unitary Fermi gas near the superfluid phase transition,''
\href{https://doi.org/10.1103/PhysRevLett.115.020401}
{Phys.\ Rev.\ Lett.\ \textbf{115}, 020401 (2015)}.
%arXiv:1410.4835 [cond-mat.quant-gas]

\bibitem{Kovtun:2005}
P.~K.~Kovtun, D.~T.~Son, and A.~O.~Starinets,
``Viscosity in strongly interacting quantum field theories from black hole physics,''
\href{https://doi.org/10.1103/PhysRevLett.94.111601}
{Phys.\ Rev.\ Lett.\ \textbf{94}, 111601 (2005)}.
%arXiv:hep-th/0405231

\bibitem{Son:2007}
D.~T.~Son,
``Vanishing bulk viscosities and conformal invariance of the unitary Fermi gas,''
\href{https://doi.org/10.1103/PhysRevLett.98.020604}
{Phys.\ Rev.\ Lett.\ \textbf{98}, 020604 (2007)}.
%arXiv:cond-mat/0511721 [cond-mat.other]

\bibitem{Elliott:2014a}
E.~Elliott, J.~A.~Joseph, and J.~E.~Thomas,
``Observation of conformal symmetry breaking and scale invariance in expanding Fermi gases,''
\href{https://doi.org/10.1103/PhysRevLett.112.040405}
{Phys.\ Rev.\ Lett.\ \textbf{112}, 040405 (2014)}.
%arXiv:1308.3162 [cond-mat.quant-gas]

\bibitem{Baird:2019}
L.~Baird, X~.Wang, S.~Roof, and J.~E.~Thomas,
``Measuring the hydrodynamic linear response of a unitary Fermi gas,''
\href{https://doi.org/10.1103/PhysRevLett.123.160402}
{Phys.\ Rev.\ Lett.\ \textbf{123}, 160402 (2019)}.
%arXiv:1906.11179 [cond-mat.quant-gas]

\bibitem{Patel:2020}
P.~B.~Patel, Z.~Yan, B.~Mukherjee, R.~J.~Fletcher, J.~Struck, and M.~W.~Zwierlein,
``Universal sound diffusion in a strongly interacting Fermi gas,''
\href{https://doi.org/10.1126/science.aaz5756}
{Science \textbf{370}, 1222-1226 (2020)}.
%arXiv:1909.02555 [cond-mat.quant-gas]

\bibitem{Vogt:2012}
E.~Vogt, M.~Feld, B.~Fr\"ohlich, D.~Pertot, M.~Koschorreck, and M.~K\"ohl,
``Scale invariance and viscosity of a two-dimensional Fermi gas,''
\href{https://doi.org/10.1103/PhysRevLett.108.070404}
{Phys.\ Rev.\ Lett.\ \textbf{108}, 070404 (2012)}.
%arXiv:1111.1173 [cond-mat.quant-gas]

\bibitem{Massignan:2005}
P.~Massignan, G.~M.~Bruun, and H.~Smith,
``Viscous relaxation and collective oscillations in a trapped Fermi gas near the unitarity limit,''
\href{https://doi.org/10.1103/PhysRevA.71.033607}
{Phys.\ Rev.\ A \textbf{71}, 033607 (2005)}.
%arXiv:cond-mat/0409660 [cond-mat.stat-mech]

\bibitem{Bruun:2005}
G.~M.~Bruun and H.~Smith,
``Viscosity and thermal relaxation for a resonantly interacting Fermi gas,''
\href{https://doi.org/10.1103/PhysRevA.72.043605}
{Phys.\ Rev.\ A \textbf{72}, 043605 (2005)}.
%arXiv:cond-mat/0504734 [cond-mat.stat-mech]

\bibitem{Braby:2010}
M.~Braby, J.~Chao, and T.~Sch\"afer,
``Thermal conductivity and sound attenuation in dilute atomic Fermi gases,''
\href{https://doi.org/10.1103/PhysRevA.82.033619}
{Phys.\ Rev.\ A \textbf{82}, 033619 (2010)}.
%arXiv:1003.2601 [cond-mat.quant-gas]

\bibitem{Bruun:2012}
G.~M.~Bruun,
``Shear viscosity and spin-diffusion coefficient of a two-dimensional Fermi gas,''
\href{https://doi.org/10.1103/PhysRevA.85.013636}
{Phys.\ Rev.\ A \textbf{85}, 013636 (2012)}.
%arXiv:1112.2395 [cond-mat.quant-gas]

\bibitem{Schafer:2012}
T.~Sch\"afer,
``Shear viscosity and damping of collective modes in a two-dimensional Fermi gas,''
\href{https://doi.org/10.1103/PhysRevA.85.033623}
{Phys.\ Rev.\ A \textbf{85}, 033623 (2012)}.
%arXiv:1111.7242 [cond-mat.quant-gas]

\bibitem{Dusling:2013}
K.~Dusling and T.~Sch\"afer,
``Bulk viscosity and conformal symmetry breaking in the dilute Fermi gas near unitarity,''
\href{https://doi.org/10.1103/PhysRevLett.111.120603}
{Phys.\ Rev.\ Lett.\ \textbf{111}, 120603 (2013)}.
%arXiv:1305.4688 [cond-mat.quant-gas]

\bibitem{Chafin:2013}
C.~Chafin and T.~Sch\"afer,
``Scale breaking and fluid dynamics in a dilute two-dimensional Fermi gas,''
\href{https://doi.org/10.1103/PhysRevA.88.043636}
{Phys.\ Rev.\ A \textbf{88}, 043636 (2013)}.
%arXiv:1308.2004 [cond-mat.quant-gas]

\bibitem{Enss:2011}
T.~Enss, R.~Haussmann, and W.~Zwerger,
``Viscosity and scale invariance in the unitary Fermi gas,''
\href{https://doi.org/10.1016/j.aop.2010.10.002}
{Ann.\ Phys.\ (NY) \textbf{326}, 770-796 (2011)}.
%arXiv:1008.0007 [cond-mat.quant-gas]

\bibitem{Nishida:2019}
Y.~Nishida,
``Viscosity spectral functions of resonating fermions in the quantum virial expansion,''
\href{https://doi.org/10.1016/j.aop.2019.167949}
{Ann.\ Phys.\ (NY) \textbf{410}, 167949 (2019)}.
%arXiv:1904.12832 [cond-mat.quant-gas]

\bibitem{Hofmann:2020}
J.~Hofmann,
``High-temperature expansion of the viscosity in interacting quantum gases,''
\href{https://doi.org/10.1103/PhysRevA.101.013620}
{Phys.\ Rev.\ A \textbf{101}, 013620 (2020)}.
%arXiv:1905.05133 [cond-mat.quant-gas]

\bibitem{Enss:2019}
T.~Enss,
``Bulk viscosity and contact correlations in attractive Fermi gases,''
\href{https://doi.org/10.1103/PhysRevLett.123.205301}
{Phys.\ Rev.\ Lett.\ \textbf{123}, 205301 (2019)}.
%arXiv:1904.12772 [cond-mat.quant-gas]

\bibitem{Frank:2020}
B.~Frank, W.~Zwerger, and T.~Enss,
``Quantum critical thermal transport in the unitary Fermi gas,''
\href{https://doi.org/10.1103/PhysRevResearch.2.023301}
{Phys.\ Rev.\ Res.\ \textbf{2}, 023301 (2020)}.
%arXiv:2003.10338 [cond-mat.quant-gas]

\bibitem{Fujii:2020}
K.~Fujii and Y.~Nishida,
``Bulk viscosity of resonating fermions revisited: Kubo formula, sum rule, and the dimer and high-temperature limits,''
\href{https://doi.org/10.1103/PhysRevA.102.023310}
{Phys.\ Rev.\ A \textbf{102}, 023310 (2020)}.
%arXiv:2004.12154 [cond-mat.quant-gas]

\bibitem{Eliashberg:1962}
G.~M.~Eliashberg,
``Transport equation for a degenerate system of Fermi particles,''
\href{http://www.jetp.ac.ru/cgi-bin/e/index/e/14/4/p886?a=list}
{Sov.\ Phys.\ JETP \textbf{14}, 886-892 (1962)}.

\bibitem{Jeon:1995}
S.~Jeon,
``Hydrodynamic transport coefficients in relativistic scalar field theory,''
\href{https://doi.org/10.1103/PhysRevD.52.3591}
{Phys.\ Rev.\ D \textbf{52}, 3591-3642 (1995)}.
%arXiv:hep-ph/9409250

\bibitem{Jeon:1996}
S.~Jeon and L.~G.~Yaffe,
``From quantum field theory to hydrodynamics: Transport coefficients and effective kinetic theory,''
\href{https://doi.org/10.1103/PhysRevD.53.5799}
{Phys.\ Rev.\ D \textbf{53}, 5799-5809 (1996)}.
%arXiv:hep-ph/9512263

\bibitem{Hidaka:2011}
Y.~Hidaka and T.~Kunihiro,
``Renormalized linear kinetic theory as derived from quantum field theory: A novel diagrammatic method for computing transport coefficients,''
\href{https://doi.org/10.1103/PhysRevD.83.076004}
{Phys.\ Rev.\ D \textbf{83}, 076004 (2011)}.
%arXiv:1009.5154 [hep-ph]

\bibitem{Liu:2013}
X.-J.~Liu,
``Virial expansion for a strongly correlated Fermi system and its application to ultracold atomic Fermi gases,''
\href{https://doi.org/10.1016/j.physrep.2012.10.004}
{Phys.\ Rep.\ \textbf{524}, 37-83 (2013)}.
%arXiv:1210.2176 [cond-mat.quant-gas]

\bibitem{Kubo:1957a}
R.~Kubo,
``Statistical-mechanical theory of irreversible processes.\ I. General theory and simple applications to magnetic and conduction problems,''
\href{https://doi.org/10.1143/JPSJ.12.570}
{J.\ Phys.\ Soc.\ Jpn.\ \textbf{12}, 570-586 (1957)}.

\bibitem{Kubo:1957b}
R.~Kubo, M.~Yokota, and S.~Nakajima,
``Statistical-mechanical theory of irreversible processes.\ II. Response to thermal disturbance,''
\href{https://doi.org/10.1143/JPSJ.12.1203}
{J.\ Phys.\ Soc.\ Jpn.\ \textbf{12}, 1203-1211 (1957)}.

\bibitem{Mori:1962}
H.~Mori,
``Collective motion of particles at finite temperatures,''
\href{https://doi.org/10.1143/PTP.28.763}
{Prog.\ Theor.\ Phys.\ \textbf{28}, 763-783 (1962)}.

\bibitem{Kadanoff:1963}
L.~P.~Kadanoff and P.~C.~Martin,
``Hydrodynamic equations and correlation functions,''
\href{https://doi.org/10.1016/0003-4916(63)90078-2}
{Ann.\ Phys.\ \textbf{24}, 419-469 (1963)}.

\bibitem{Luttinger:1964}
J.~M.~Luttinger,
``Theory of thermal transport coefficients,''
\href{https://doi.org/10.1103/PhysRev.135.A1505}
{Phys.\ Rev.\ \textbf{135}, A1505-A1514 (1964)}.

\bibitem{Altland-Simons}
See, for example, A.~Altland and B.~D.~Simons,
\textit{Condensed Matter Field Theory,} 2nd ed.\
(Cambridge University Press, Cambridge, U.K., 2010).

\bibitem{Fujii:2018}
K.~Fujii and Y.~Nishida,
``Hydrodynamics with spacetime-dependent scattering length,''
\href{https://doi.org/10.1103/PhysRevA.98.063634}
{Phys.\ Rev.\ A \textbf{98}, 063634 (2018)}.
%arXiv:1807.07983 [cond-mat.quant-gas]

\bibitem{Martinez:2017}
M.~Martinez and T.~Sch\"afer,
``Hydrodynamic tails and a fluctuation bound on the bulk viscosity,''
\href{https://doi.org/10.1103/PhysRevA.96.063607}
{Phys.\ Rev.\ A \textbf{96}, 063607 (2017)}.
%arXiv:1708.01548 [cond-mat.quant-gas]

\bibitem{Leyronas:2011}
X.~Leyronas,
``Virial expansion with Feynman diagrams,''
\href{https://doi.org/10.1103/PhysRevA.84.053633}
{Phys.\ Rev.\ A \textbf{84}, 053633 (2011)}.
%arXiv:1109.4794 [cond-mat.quant-gas]

\bibitem{Lifshitz-Pitaevskii}
See, for example, E.~M.~Lifshitz and L.~P.~Pitaevskii,
\textit{Physical Kinetics}
(Pergamon Press, Oxford, U.K., 1981).

\end{thebibliography}
\end{document}